\documentclass[%
 superscriptaddress,
 reprint,
 amsmath,amssymb,amsthm,mathrsfs,
 aps,
prb,
]{revtex4-1}

\usepackage{graphicx}
\usepackage{dcolumn}
\usepackage{bm}

\usepackage[ 
margin=0.75in,
]{geometry}

\def\p4mmm{$P4/mmm$}
\def\btao{BaTi$_{2}$As$_{2}$O}
\def\btso{BaTi$_{2}$Sb$_{2}$O}

\def\btpo{BaTi$_{2}Pn_{2}$O}
\def\btbo{BaTi$_{2}$Bi$_{2}$O}

\usepackage[usenames]{color}

\def\sjba#1{}
\def\bafa#1{}

\begin{document}

\preprint{}

\title{
Intra-unit-cell nematic charge order in the titanium-oxypnictide family of superconductors
}

\author{Benjamin A. Frandsen$^*$}
\affiliation{%
 Department of Physics, Columbia University, New York, NY 10027, USA.
}%

\author{Emil S. Bozin}
\thanks{These authors contributed equally to this work.}
\affiliation{%
 Condensed Matter Physics and Materials Science Department, Brookhaven National Laboratory, Upton, NY 11973, USA.
}%

\author{Hefei Hu}
\affiliation{%
 Condensed Matter Physics and Materials Science Department, Brookhaven National Laboratory, Upton, NY 11973, USA.
}%

\author{Yimei Zhu}
\affiliation{%
 Condensed Matter Physics and Materials Science Department, Brookhaven National Laboratory, Upton, NY 11973, USA.
}%

\author{Yasumasa Nozaki}
\affiliation{%
 Department of Energy and Hydrocarbon Chemistry, Graduate School of Engineering, Kyoto University, Nishikyo, Kyoto 615-8510, Japan.
}%

\author{Hiroshi Kageyama}
\affiliation{%
 Department of Energy and Hydrocarbon Chemistry, Graduate School of Engineering, Kyoto University, Nishikyo, Kyoto 615-8510, Japan.
}%

\author{Yasutomo J. Uemura}
\affiliation{%
 Department of Physics, Columbia University, New York, NY 10027, USA.
}%

\author{Wei-Guo Yin}
\affiliation{%
 Condensed Matter Physics and Materials Science Department, Brookhaven National Laboratory, Upton, NY 11973, USA.
}%

\author{Simon J. L. Billinge}
\email{sb2896@columbia.edu}
\affiliation{%
 Condensed Matter Physics and Materials Science Department, Brookhaven National Laboratory, Upton, NY 11973, USA.
}%
\affiliation{%
 Department of Applied Physics and Applied Mathematics, Columbia University, New York, NY 10027, USA.
}%

\date{\today}

\begin{abstract}

Understanding the role played by broken symmetry states such as charge, spin, and orbital orders in the mechanism of emergent properties such as high-temperature superconductivity is a major current topic in materials research. That the order may be within one unit cell, such as nematic, was only recently considered theoretically, but its observation in the iron-pnictide and doped cuprate superconductors places it at the forefront of current research. Here we show that the recently discovered \btso\ superconductor and its parent compound \btao\ form a symmetry-breaking nematic ground state that can be naturally explained as an intra-unit-cell nematic charge order with $d$-wave symmetry, pointing to the ubiquity of the phenomenon. These findings, together with the key structural features in these materials being intermediate between the cuprate and iron-pnictide high-temperature superconducting materials, render the titanium oxypnictides an important new material system to understand the nature of nematic order and its relationship to superconductivity.

\end{abstract}

\maketitle

\section{Introduction}
Rather than being an anomalous side-effect in one or two cuprate systems, broken-symmetry states are now thought to be widespread in strongly correlated electron systems and other complex materials. Extensive study of the manganites~\cite{dagot;b;npsacm03, dagot;s05}, cuprates~\cite{milli;s00, dasil;s14}, iron pnictides~\cite{ferna;np14}, and a variety of other systems has made it increasingly evident that local and global symmetry-breaking in the charge, orbital, lattice, and spin degrees of freedom are associated with the appearance of emergent phenomena such as colossal magnetoresistance and high-temperature superconductivity (HTSC), but the exact relationship is not understood. Historically, the study of such broken-symmetry states has been very challenging. Taking the cuprates as an example, eight years elapsed from the initial discovery of superconductivity to the first observation of symmetry-broken charge order (stripes) in one system~\cite{tranq;n95}, another seven years passed before hints were found in others~\cite{hoffm;s02, versh;s04}, and only within the last three years has charge order begun to emerge as a possibly ubiquitous feature of the cuprates~\cite{chang;np12, comin;s14}.

Several possibilities arise when considering the symmetries that can be broken by these states. Most charge/spin-density-waves (C/SDWs) break the translational symmetry of the lattice, folding the Brillouin zone and resulting in superlattice diffraction peaks. On the other hand, orbital ordering, where charge transfers between orbitals centered at the same site, can break the metric rotational symmetry without lowering the translational symmetry. Examples are charge-nematic~\cite{lawle;n10} and loop-current~\cite{varma;prb97, li;n10} orders in the doped cuprates. In this context, nematic order is defined as one that breaks the rotational point group symmetry while preserving the lattice translational symmetry. The fact that nematic symmetry-broken states have recently been discovered experimentally in both the cuprate~\cite{lawle;n10} and iron-based~\cite{chuan;s10} superconductors raises the importance and relevance of this observation to HTSC. It is therefore critically important to understand the role and ubiquity of symmetry breaking, including intra-unit-cell nematicity, to the superconducting phenomenon.

Standard theoretical treatments of HTSC, such as the effective single-band $t-J$ model, have typically ignored the possibility of intra-unit-cell orders~\cite{lee;rmp06}. When multiple atoms per unit cell are explicitly included in the theory, qualitatively different ground-state solutions may be found~\cite{ferna;prb12}, underscoring the subtlety and importance of accounting correctly for this phenomenon.  Hence, finding related but distinct systems that exhibit this phenomenon is expected to shed new light on this critical question.

Very recently, superconductivity was discovered~\cite{yajim;jpsj13i, yajim;jpsj13ii, zhai;prb13} in titanium-oxypnictide compounds such as $A$Ti$_2 Pn_2$O ($A = $Na$_2$, Ba, (SrF)$_2$, (SmO)$_2$; $Pn = $As, Sb, Bi), which are close structural and chemical cousins to the cuprates and iron-pnictides~\cite{ozawa;stam08, wang;jpcm10, liu;cm10, johre;zk11, yajim;jpsj12}. In particular, in isovalent BaTi$_2$(Sb$_{1-x}$Bi$_x$)$_2$O and aliovalent Ba$_{1-x}$Na$_x$Ti$_2$Sb$_2$O,~\cite{doan;jacs12} muon spin rotation and heat capacity measurements point to fully-gapped $s$-wave superconductivity~\cite{nozak;prb13, vonRo;prb13, gooch;prb13}. Interestingly, a number of compounds in this family also show strong anomalies in resistivity and/or magnetic susceptibility that are thought to be signatures of symmetry-breaking charge- or spin-ordered ground states~\cite{axtel;jssc97, liu;prb09, wang;jpcm10, doan;jacs12, yajim;jpsj12}, suggesting that these materials are excellent candidates for studying the interplay between broken-symmetry states and superconductivity. In light of these strong transport anomalies, it is then quite surprising that subsequent experiments have failed to uncover any direct evidence for a conventional spin- or charge-density wave ground state~\cite{kitag;prb13, nozak;prb13, vonRo;prb13}, leaving open the question of whether these materials do possess symmetry-broken ground states. 

Here we show that superconducting \btso , and its non-superconducting parent compound \btao , do indeed undergo a tetragonal-orthorhombic phase transition, corresponding to a $C_4-C_2$ symmetry lowering, that occurs at the temperature of the transport anomaly. On the other hand, high-sensitivity electron diffraction measurements failed to detect any superlattice peaks in the bulk at low temperature, indicating that this transition does not break translational symmetry. The low-temperature phase therefore constitutes a nematic state. In light of the pronounced upturn in resistivity accompanying this nematic transition, together with the absence of any ordered SDW~\cite{nozak;prb13}, we attribute the nematicity to an intra-unit-cell charge order with $d$-wave symmetry by charge transfer between neighboring Ti sites---similar to that between neighboring oxygen sites in cuprate superconductors---and find that it naturally explains the temperature dependence of the lattice constants. These results establish this family of materials as another playground for studying symmetry-breaking electronic phases and their relationship to superconductivity.

\section{Results}

\subsection{Structural and electronic properties of \btpo}
\begin{figure*}
\includegraphics[width=160mm]{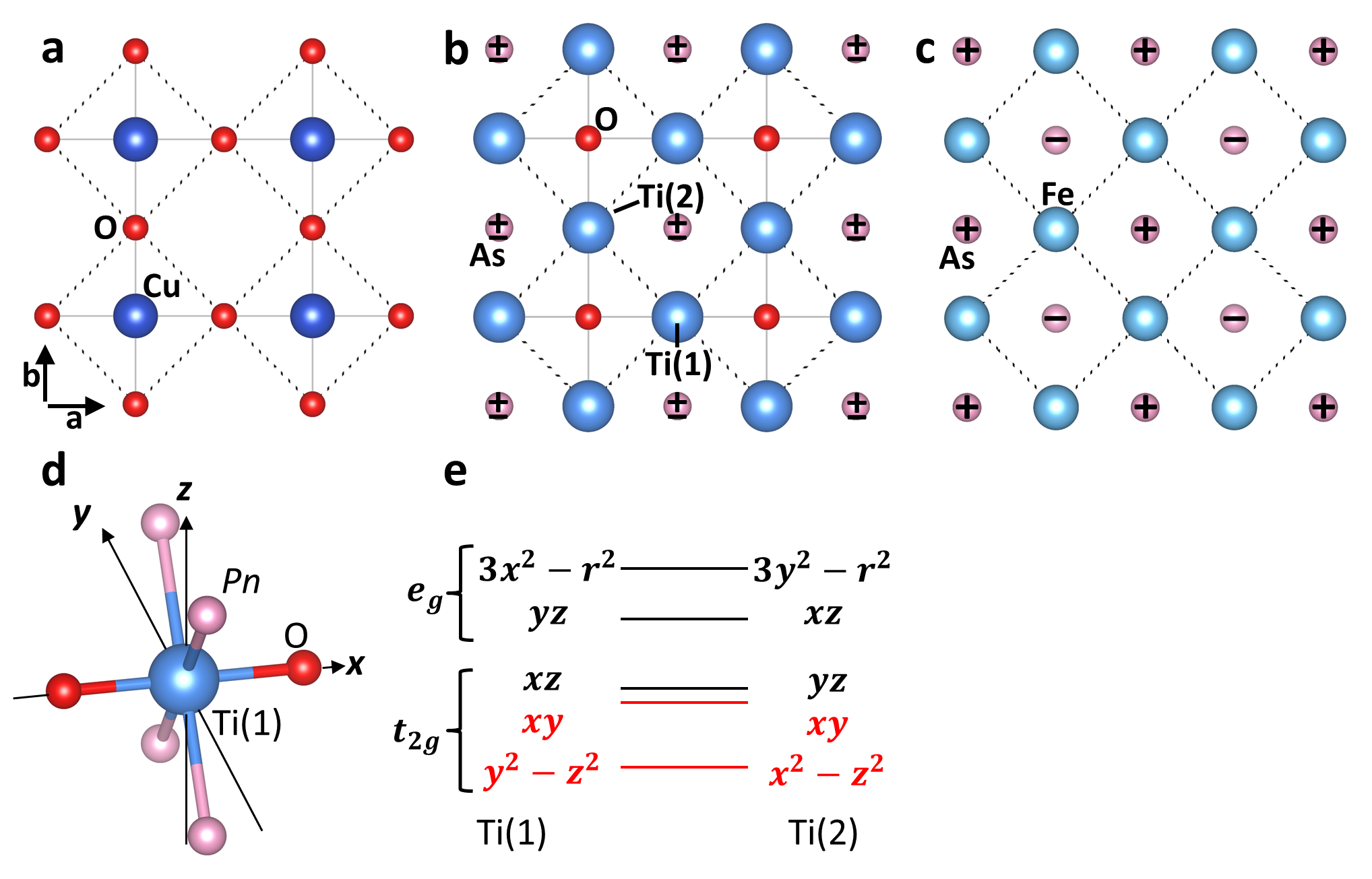}
\caption{\label{fig:planes} Planar geometries of cuprate, titanium-oxypnictide, and iron-pnictide superconducting families.  (a-c) Planar motifs in cuprates, titanium oxypnictides, and iron pnictides, respectively. Solid gray lines show the metal-anion square net, and dotted black lines show the square net of second-nearest-neighbor atoms. The + and - signs in (b) and (c) denote As atoms that are above and below the plane, respectively. (d) TiO$_2 Pn_4$ octahedral motif found in \btpo. (e) Schematics of the Ti 3$d$-orbital energy levels for the two distinct Ti sites labeled in (b), Ti(1) and Ti(2). The two lowest-lying orbitals marked in red form two bands occupied by one electron per Ti.}

\end{figure*}

\begin{figure}
\includegraphics[width=80mm]{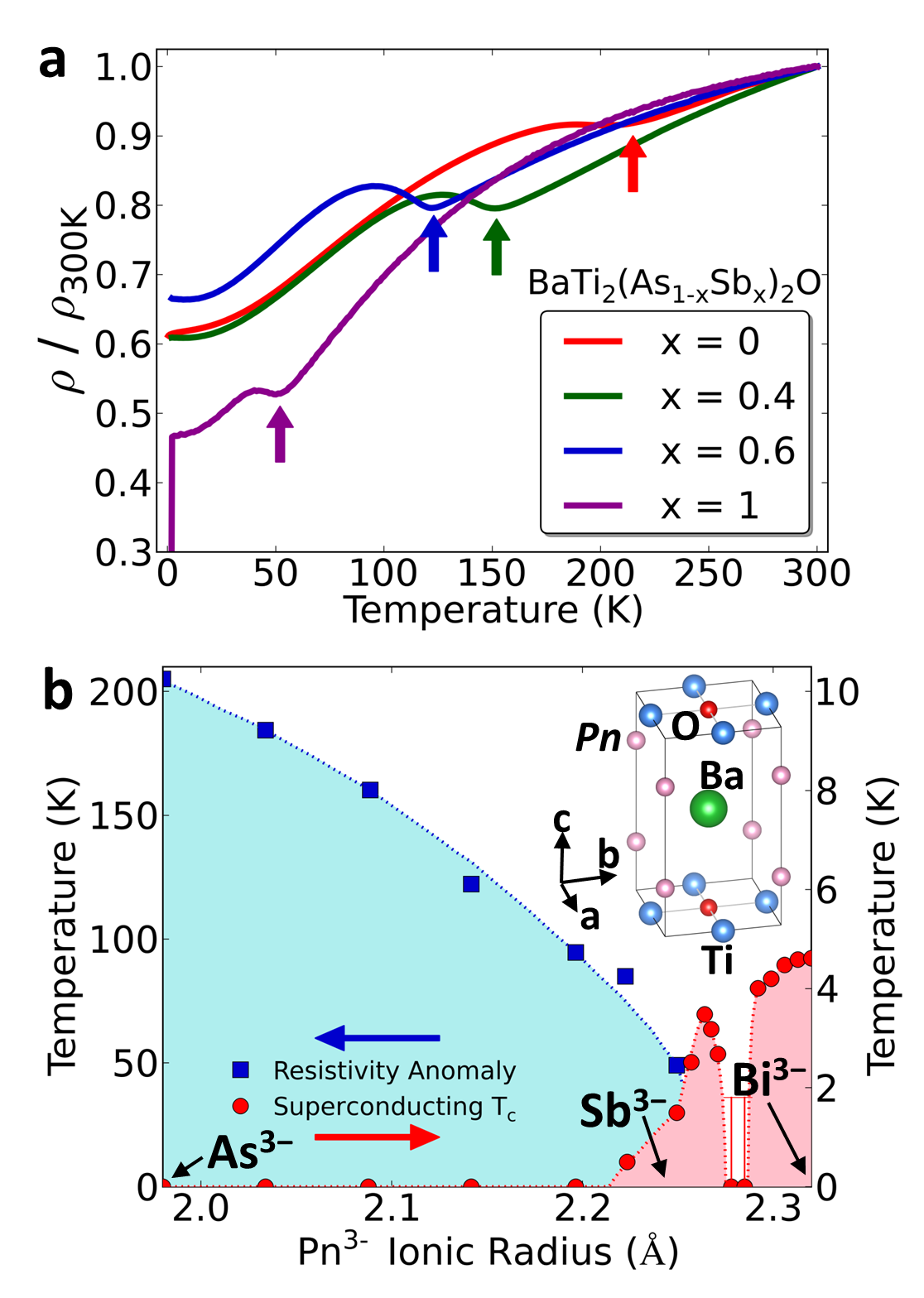}
\caption{\label{fig:resPD} Transport characteristics and phase diagram of \btpo\ with $Pn$=As, Sb, Bi. (a) Electrical resistivity of BaTi$_2$(As$_{1-x}$Sb$_x$)$_2$O, normalized by the room-temperature resistivity. Arrows indicate the anomaly discussed in the text. (b) Phase diagram of \btpo\ shown as a function of $Pn^{3-}$ ionic diameter. Broken lines are guides to the eye. The error bars accompanying the red circles arise from the instrumental low-temperature limit of 1.8~K. Inset: tetragonal crystal structure of \btpo.}
\end{figure}

The basic structural unit of \btpo\ is a planar square net of titanium and oxygen, in analogy with the cuprates [Fig.~\ref{fig:planes}(a) and (b)], with the crucial difference that the positions of the metal and oxygen ions are switched between the structures [the complete titanate structure is shown as an inset in Fig.~\ref{fig:resPD}(b)].
In Fig.~\ref{fig:planes}, the square net is shown by solid lines along the nearest neighbor bonds, with dashed lines showing the net joining second neighbor ions.  This second-nearest-neighbor square net connects oxygen ions in the cuprates, but metal ions in the titanate compounds and also in the iron-based superconductors [Fig.~\ref{fig:planes}(c)]. Thus, in terms of chemistry and structure, the titanate compounds bridge between the ferrous and cuprate superconductors.  The Ti 3$d$ orbitals are occupied by one electron per Ti atom, which is found to reside in a nominally 1/4-filled band formed via hybridization of the $d_{xy}$ and $d_{y^2 - z^2}$/$d_{x^2 - z^2}$ orbitals for the Ti(1)/Ti(2) ions (defined in Fig.~\ref{fig:planes}(b)). The local geometry of the Ti(1) site is shown in Fig.~\ref{fig:planes}(d) and the arrangement of the $d$-energy levels in Fig.~\ref{fig:planes}(e).

Furthermore, the phase diagram [Fig.~\ref{fig:resPD}(b)] is highly reminiscent of the cuprates and iron-based superconductors, with superconductivity appearing on doping and transport behavior that is strongly suggestive of a competing electronic transition such as the formation of a CDW or SDW~\cite{wang;jpcm10, yajim;jpsj12}.
The transport is metallic at high temperature~\cite{wang;jpcm10}, with a positive resistivity slope vs. temperature [Fig.~\ref{fig:resPD}(a)]. However, on cooling a pronounced upturn in the resistivity is found for all $x$ in the solid solution BaTi$_2$As$_{1-x}$Sb$_x$O. The feature occurs at a temperature $T_a$ that decreases monotonically from 200~K for $x=0$ to 50~K for $x=1$, with superconductivity appearing below approximately 1~K for the antimony endmember and increasing to 5~K for \btbo.~\cite{yajim;jpsj12, zhai;prb13}
Anomalies in the magnetic susceptibility and specific heat are also observed~\cite{wang;jpcm10} at $T_a$.

Density functional theory (DFT) calculations for \btso\ predicted an instability towards a bicollinear SDW formation~\cite{singh;njp12, wang;jap13} or a commensurate CDW ground state driven by an unstable phonon mode that doubles the unit cell by distorting the Ti squares and preserves the tetragonal symmetry~\cite{subed;prb13}. The possibility of SDW formation in BaTi$_{2}$(As,Sb)$_{2}$O, either commensurate or incommensurate, was subsequently ruled out by muon spin relaxation and $^{121/123}$Sb nuclear resonance measurements, which show conclusively that no magnetic order develops at any temperature probed~\cite{kitag;prb13,nozak;prb13,vonRo;prb13}. On the other hand, a conventional CDW should be evident through an associated structural distortion. However, initial electron and neutron diffraction studies on the Sb endmember~\cite{nozak;prb13} found no broken symmetry or any signature of superlattice formation at low temperatures, nor was a CDW gap formation observed in angle-resolved photoemission measurements of the nested Fermi surfaces (although a slight depression of the density of states at other momenta was found to correlate in temperature with the resistivity anomaly~\cite{xu;prb14}). The possibility of an incommensurate CDW was also ruled out by $^{121/123}$Sb nuclear resonance measurements~\cite{kitag;prb13}.

\subsection{Neutron powder diffraction measurements}
\begin{figure}
\includegraphics[width=80mm]{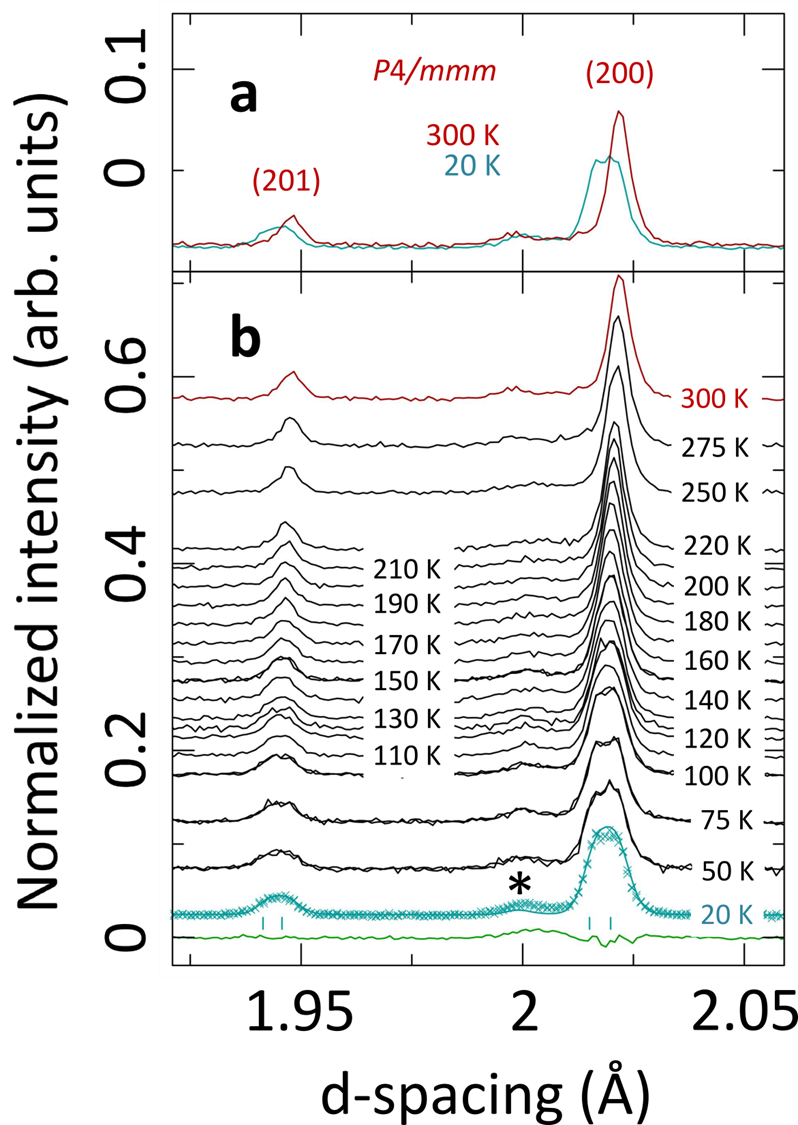}
\caption{\label{fig:btao-dspacing} Temperature evolution of \btao\ neutron diffraction pattern. (a) Comparison of normalized intensities of 300~K (red) and 20~K (blue) data around (200) and (201) reflections in \p4mmm setting. (b) Waterfall plot across the temperature range studied. The 20~K data (blue symbols) are shown with a fit (blue solid line) of the refined $Pmmm$ model. The corresponding difference curve is shown as the green solid line below. The asterisk marks the (200)/(020) reflections of the BaTiO$_3$ impurity phase.}
\end{figure}

In the absence of evidence for a long-range ordered CDW, we undertook a neutron diffraction and total scattering measurement on \btso\ with a dense set of temperature points to search for evidence for a possible short-range ordered CDW~\cite{abeyk;prl13,bozin;prl00}. We also extended the investigation to the previously unstudied \btao\ endmember. Unexpectedly, we found a long-range structural phase transition at $T_a$. Room temperature measurements of \btao\ confirm the tetragonal \p4mmm space group symmetry previously reported from x-ray diffraction~\cite{wang;jpcm10}. However, on cooling through $T_a$ we observe a distinct splitting of the (200)/(020) and (201)/(021) Bragg peaks, representing the first observation of a symmetry lowering at the same temperature as the resistivity anomaly in \btao. This is shown in Fig.~\ref{fig:btao-dspacing}, which compares the high- and low-temperature Bragg peaks in panel (a) and displays their temperature evolution in panel (b). 
The (200) peak at 2.02~\AA\ begins to broaden below 200~K, coinciding with $T_a$, and appears to split at the lowest temperatures. Similarly, the (201) peak at 1.94~\AA\ displays apparent splitting as the temperature is lowered. These observations demonstrate that \btao\ undergoes a long-range-ordered structural phase change at $T_a$, lowering its symmetry from tetragonal to orthorhombic. \sjba{oh, I see what this is about.  I suggest that you rewrite this so it is less confusing.  Languange like "splitting of the 200 and 020 peaks" is very common and is what people will think you mean if you use that language.  a (200) peak can't split, though a <200> peak can....but just be clear.}

\begin{figure}
\includegraphics[width=80mm]{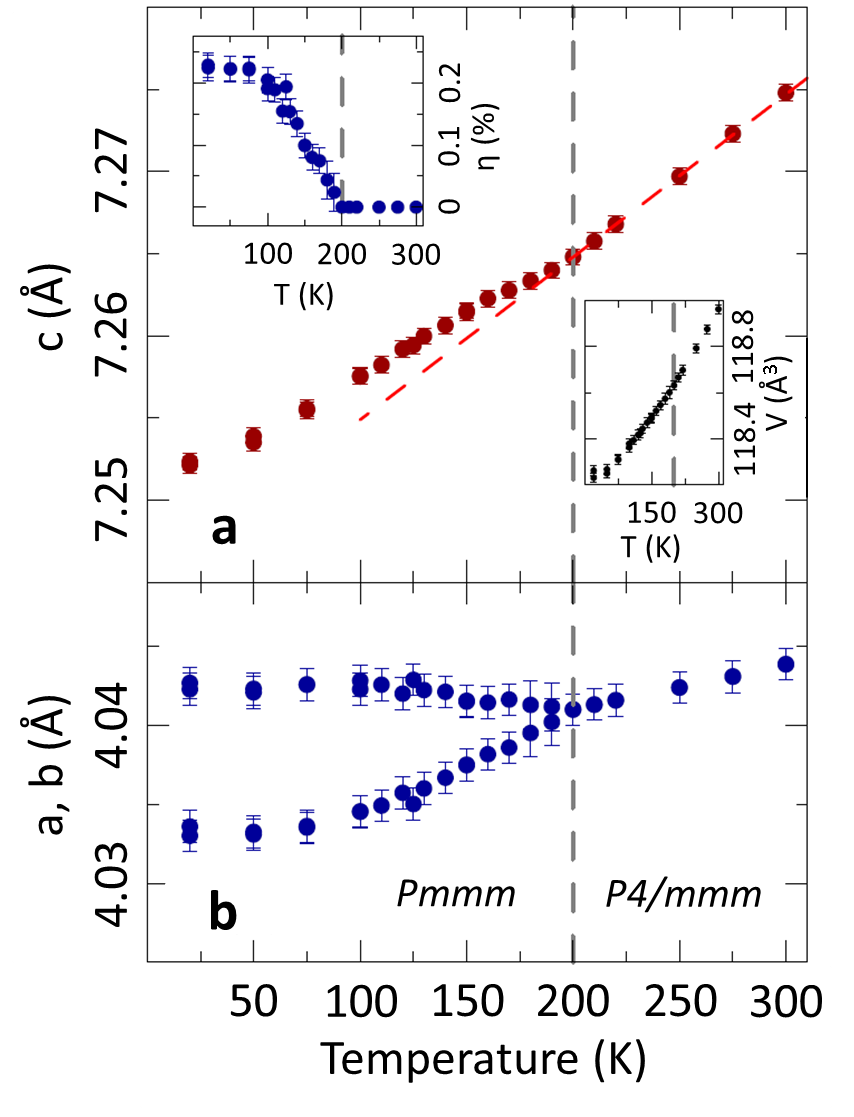}
\caption{\label{fig:fitResultBTAO} Temperature evolution of \btao\ structural parameters. (a) Lattice parameter $c$ (red). (b) Lattice parameters $a$, $b$ (blue). Insets: orthorhombicity $\eta = 2 \times (a-b)/(a+b)$ (left) and unit cell volume (right). Vertical dashed gray line indicates transition temperature. Dashed red line is a guide for the eyes. Error bars represent the estimated standard deviation of the corresponding refined parameter. }
\end{figure}

To investigate the structural transition in greater detail, we performed LeBail~\cite{lebail;mrb87} refinements at all temperatures. We used the parent \p4mmm model for $T \ge 200$~K. The simplest possible symmetry-breaking distortion mode of the parent \p4mmm structure consistent with the observed peak splitting is a mode that breaks the degeneracy of the $a$- and $b$-axes without otherwise shifting atoms within the unit cell, resulting in a space-group symmetry of $Pmmm$, which was used for $T < 200$~K. We also explored other candidate orthorhombic structures with lower symmetry, but since these structures do not improve the fit quality within the resolution limitations of the current data, $Pmmm$ is the most appropriate choice. We display the results of these refinements in Fig.~\ref{fig:fitResultBTAO}.
As seen in panel (b), the tetragonal $a$-axis clearly splits below $T \sim 200$~K, with a maximum orthorhombic splitting of approximately 0.01~\AA. The orthorhombicity parameter $\eta = 2 \times (a-b)/(a+b)$ is shown in the inset of panel (a), indicating a maximum orthorhombicity of $\sim$~0.22\%. Panel (a) also displays the temperature dependence of the $c$-axis parameter, which exhibits an upturn below the structural transition deviating from the linear thermal contraction trend seen for $T>200$~K. This same type of $c$-axis response also accompanies long-range ordered stripe formation in the nickelates~\cite{abeyk;prl13}. Additional details regarding Rietveld refinement of \btao\ can be found in Supplementary Note 1. The superconducting \btso\ shows qualitatively the same behavior, albeit with an amplitude decreased by a factor of approximately 5, with an orthorhombic splitting (0.05\%) and a small but observable $c$-axis upturn appearing on cooling through $T_a= $~50~K. The undistorted \p4mmm model can be used with moderate success at all temperatures, but the $Pmmm$ model yields a better fit below 50~K (see Supplementary Note 2). Pair distribution function analysis is consistent with these observations for both compounds and can be found in Supplementary Note 3.

These results offer compelling evidence that the observed structural response is intimately related to the transport anomaly and may be driven by a broken symmetry of the electronic system forming at that temperature. The small distortion amplitude in \btso\ explains why this long-range structural phase change escaped notice in previous neutron diffraction measurements~\cite{nozak;prb13}.

\subsection{Electron diffraction measurements}
\begin{figure}
\includegraphics[width=80mm]{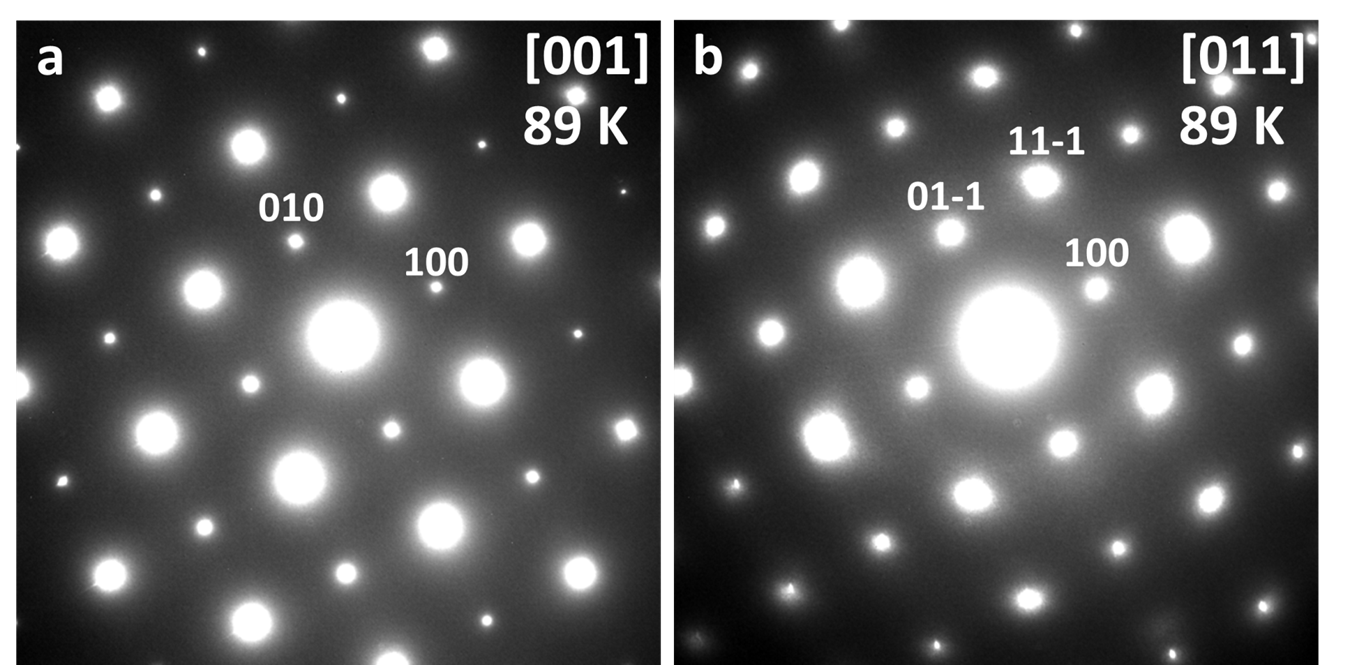}
\caption{\label{fig:ED} Electron diffraction patterns of \btao. (a) Diffraction pattern with the incident beam along the [001] and (b) along the [011] directions. No superlattice peaks are observed at low temperature even after heavy overexposure.}
\end{figure}

Since CDW formation is implicated, we made a special effort to search for the appearance of weak superlattice peaks associated with a finite CDW wavevector in electron-diffraction (ED) patterns below $T_a$. The original study on the antimony endmember failed to observe superlattice peaks~\cite{nozak;prb13}. Here we concentrated on the arsenic endmember where the structural distortion is five times larger, and the ED patterns were heavily overexposed to search for any weak response at intensities close to background. Despite these efforts, the ED patterns taken along the [001] and [011] directions revealed no superlattice peaks in the bulk at low temperature, as shown in Fig.~\ref{fig:ED}.
However, in a very small fraction of the sample in the immediate vicinity of grain boundaries, weak superlattice peaks with $\mathbf{Q} = (1/2,0,0)$ are observed at low temperature. This non-bulk behaviour is explored further in Supplementary Note 4.

\section{Discussion}
A picture emerges of a $C_4-C_2$ symmetry breaking occurring with an accompanying strong upturn in resistivity, but with no corresponding CDW superlattice peaks appearing. The resistivity upturn is larger than would be expected as a passive response of the electronic system to the structural transition, borne out by a standard DFT calculation with the observed orthorhombicity parameter $\eta = 0.22\%$ which showed that merely 0.0003 electrons are transferred from Ti(1) to Ti(2). Therefore, in common with earlier discussion~\cite{wang;jpcm10,yajim;jpsj13ii}, we propose that the structural transition is a response to an instability of the electronic system. The earlier muon spin relaxation results~\cite{nozak;prb13,vonRo;prb13}, together with the ED measurements, allow us to rule out the existing proposals of SDW formation~\cite{singh;njp12,wang;jap13} or a phonon-driven CDW~\cite{subed;prb13}. Instead, an intra-unit-cell charge-nematic electronic symmetry breaking is implicated, similar to that proposed for doped cuprates~\cite{fujit;pnas14}.

In the current case, a charge redistribution between the on-site orbital states at the Fermi level, $d_{x^2 - z^2}$/$d_{y^2 - z^2}$ to $d_{xy}$, does not break the rotational symmetry so can be ruled out. Instead, a simple but novel intra-unit-cell charge order (IUC-CO) naturally explains the observed phenomenology.  A transfer of charge from Ti(1) to Ti(2) [see Fig.~\ref{fig:planes}(b)] lowers the rotational symmetry of the Ti$_2$O plaquette locally from $C_4$ to $C_2$, with the effect on the overall lattice symmetry depending on the ordering pattern of the distinct Ti ions in neighboring unit cells. Repeating the symmetry-lowered plaquette uniformly along the $a$- and $b$-directions results in no change of the unit cell, but breaks the metric symmetry from $C_4$ in \p4mmm to $C_2$ in $Pmmm$, as observed experimentally. This arrangement of charges can accordingly be described as a nematic IUC-CO. Our data are therefore consistent with the formation of this type of charge order on cooling through $T_a$. Such a model is also consistent with the breaking of $C_4$ symmetry at the $Pn$ site that has been observed from $^{121/123}$Sb nuclear resonance measurements~\cite{kitag;prb13}.

This nematic IUC-CO is energetically favored on Coulombic grounds if the onsite 
Hubbard energy $U$ is sufficiently small, which is a reasonable assumption 
since the system is a metal rather than a Mott insulator. The small Hubbard $U$ arises from significant screening due to the solvation effect of the high polarizabilities of the As$^{3-}$ and Sb$^{3-}$ anions, which are an order of magnitude larger than that of O$^{2-}$. Moreover, the As and Sb ions reside on a lower symmetry site in the $Pmmm$ structure, which enhances the effects of their polarizabilities. This physics was proposed in an earlier study of the iron-pnictide superconductors~\cite{sawat;el09}. Returning to the present system, the transfer of a 
charge of $\delta$ from Ti(1) to Ti(2) will result in a lowering of the 
electrostatic energy, $V(1-\delta)(1+\delta)=V(1-\delta^2)$, where $V$ is 
the screened Coulomb interaction between Ti sites and is positive. 
The result is charge order with a $d$-wave symmetry~\cite{fradk;arocmp10} 
with the sign of the modulated charge density varying as $-+-+$ around the 
plaquette.  From plaquette to plaquette, the orientation of the axis of the 
distortion can be parallel or perpendicular, forming a ferro- or anti-ferro-
type ordering which would preserve or break translational symmetry, 
respectively. The former is consistent with the experimental observations 
in this material. We suggest that the rather rigid face-shared octahedral 
topology in each layer favors the uniform ferro- over the anti-ferro-
ordering. It is noteworthy that this $V$ is the counterpart of the Coulumb repulsion $V_{pp}$ between neighboring oxygen atoms in the CuO$_2$ plane, which was shown to drive IUC nematic order in nonstoichiometric doped cuprates~\cite{fisch;prb11}. Hence, the present results obtained on these stoichiometric materials (thus having less ambiguity such as disorder effects) yield insight into the origin of IUC nematic order in the cuprates. 

This nematic IUC-CO naturally explains the observed changes in the $a$- and $c$-lattice parameters. The transfer of charge from the Ti(1) $d_{y^2 - z^2}$ orbital to Ti(2) $d_{x^2 - z^2}$ results in increased electrostatic repulsion between the charge-rich $d_{x^2 - z^2}$ orbitals extending along the $a$-axis, breaking the tetragonal degeneracy of the $a$- and $b$-axes and leading to the observed orthorhombic distortion. Furthermore, a uniform stacking of the IUC-CO in each layer can also explain the response of the $c$-lattice parameter, which expands upon entering the charge-ordered state [Fig.~\ref{fig:fitResultBTAO}(a)]. This lattice expansion may be attributed in part to increased electrostatic repulsion between inter-layer Ti ions from the transferred charge.  The net energy contribution is $V'[(1+\delta)^2+(1-\delta)^2]=V'(2+2\delta ^2)$, where $V'$ is the inter-layer screened Coulomb interaction. This acts in addition to other elastic energy contributions.

The structural effects observed on cooling are much smaller in \btso\ than \btao, suggesting that the IUC-CO is relatively suppressed both in amplitude and temperature. This may be a result of the larger unit cell in the Sb compound, due to its larger Sb$^{3-}$ ionic diameter (2.25~\AA~vs. 1.98~\AA\ for As$^{3-}$)~\cite{yajim;jpsj13ii}, resulting in a smaller $V$.

To identify the microscopic driving force of this nematic instability, we present a symmetry-based zero-order analysis in which only the leading energy scales are retained. Ti atoms reside at the center of a distorted octahedron with oxygen at the apices and pnictide atoms around the equatorial plane, as shown in Fig.~\ref{fig:planes}(d) for Ti(1). The Ti $3d$-energy levels are illustrated in Fig.~\ref{fig:planes}(e), similar to the case of nearly isostructural (LaO)$_2$CoSe$_2$O.~\cite{wu;prb10} The nominal electron occupation is one electron per Ti atom, which would have been assigned to the locally lowest lying $d_{y^2 - z^2}$ and $d_{x^2 - z^2}$ orbitals on the Ti(1) and Ti(2) ions, respectively. However, the $\sigma$-bonding between the Ti(1) and Ti(2) $d_{xy}$ orbitals forms a relatively wide band that overlaps the locally lowest level. Therefore, the minimum model for the titanium oxypnictides involves the two orbitals: $d_{y^2 - z^2}$ and $d_{xy}$ on Ti(1) and $d_{x^2 - z^2}$ and $d_{xy}$  on Ti(2). Since the $d_{xy}$ orbital has the same impact on both the $a$ and $b$ direction, the $C_4-C_2$ symmetry lowering around the central oxygen atom is mainly determined by the charge imbalance between the quasi-one-dimensional $d_{y^2 - z^2}$ band on Ti(1) and $d_{x^2 - z^2}$ on Ti(2) as a result of the Stoner instability~\cite{fisch;prb11}. No doubt this mechanism will be complicated by hybridization and other issues, but this symmetry-based analysis provides the appropriate realistic starting point. Since the proposed nematic intra-unit-cell charge order elegantly explains all the observed structural effects, it is anticipated to be the electronic ground state of \btpo.

\section{Methods}

\subsection{Neutron powder diffraction}
Powder specimens of \btao\ and \btso\ were prepared via conventional solid state reaction methods. Details of the synthesis are provided in a previous study~\cite{nozak;prb13}. Time-of-flight neutron total scattering experiments were performed at the Neutron Powder Diffractometer at Los Alamos Neutron Science Center (LANSCE) at Los Alamos National Laboratory. Data were collected using a closed-cycle He refrigerator at temperatures ranging from 10 - 300~K in steps of 10~K near the structural transition and 25~K away from the transition over a wide range of momentum transfer $Q$. Le~Bail~\cite{lebail;mrb87} fits to the intensity profiles were performed with GSAS~\cite{larso;unpub94} on the EXPGUI platform~\cite{toby;jac01}. Pair distribution function (PDF) profiles were obtained by Fourier transforming the measured total scattering intensity up to a maximum momentum transfer of $Q_{max}=24$~\AA$^{-1}$~using established protocols~\cite{egami;b;utbp12,chupa;jac03} as implemented in the program PDFgetN~\cite{peter;jac00}. The Le Bail fits were used to extract lattice parameters and space group symmetry, the PDF fits to extract atomic displacement parameters. Symmetry mode analysis using the program ISODISTORT~\cite{campb;jac06} was conducted to identify candidate distorted structures. 

\subsection{Electron diffraction and DFT calculations}
Electron diffraction patterns were recorded using a JEOL ARM 200CF transmission electron microscope (TEM), operated at 200 keV, at Brookhaven National Laboratory. The TEM samples were prepared by crushing powder specimens into thin flakes transparent to the electron beam, which were supported by a lacey carbon copper grid. DFT calculations were performed within the generalized gradient approximation (GGA) implemented in the Wien2k software package~\cite{blaha;manual01}.

\textbf{Acknowledgements}

Work at Brookhaven National Laboratory was supported by the U.S. Department of Energy, Office of Basic Energy Sciences, under contract No. DE-AC02-98CH10886. Work at Columbia University was supported by the U.S. National Science Foundation (NSF) Partnership for International Research and Education (PIRE) Super-PIRE project (grant OISE-0968226). YJU also acknowledges support from NSF DMR-1105961, the Japan Atomic Energy Agency Reimei project, and the Friends of Todai Inc. The work at Kyoto University was supported by the FIRST program, Japan Society of the Promotion of Science (JSPS). Neutron scattering experiments were carried out on NPDF at LANSCE, funded by DOE Office of Basic Energy Sciences. LANL is operated by Los Alamos National Security LLC under DOE Contract No. DE-AC52-06NA25396.


\end{document}